\documentclass[twocolumn,superscriptaddress]{revtex4-2}

\usepackage{epsfig}
\usepackage{amsmath, amssymb}
\usepackage{mathtools}
\usepackage{graphicx}
\usepackage{footnote}
\usepackage{bbm}
\usepackage{bm}
\usepackage{appendix}
\usepackage{todonotes}
\usepackage{upgreek}
\DeclareMathOperator{\Tr}{Tr}
\usepackage[normalem]{ulem}
\usepackage{xcolor}

\usepackage[all]{xy}

\definecolor{nicered}{RGB}{210,50,0}
\usepackage[colorlinks,breaklinks,bookmarks=true,citecolor=blue,linkcolor=nicered,urlcolor=magenta]{hyperref}

\begin{document}

\title{Charge order in the kagome lattice Holstein model: A hybrid Monte Carlo study}

\author{Owen Bradley}
\affiliation{Department of Physics, University of California, Davis, California 95616, USA}

\author{Benjamin Cohen-Stead}
\affiliation{Department of Physics, University of California, Davis, California 95616, USA}
\affiliation{Department of Physics and Astronomy, The University of Tennessee, Knoxville, Tennessee 37996, USA}
\affiliation{Institute for Advanced Materials and Manufacturing, The University of Tennessee, Knoxville, Tennessee 37996, USA\looseness=-1}
\author{Steven Johnston}
\affiliation{Department of Physics and Astronomy, The University of Tennessee, Knoxville, Tennessee 37996, USA}
\affiliation{Institute for Advanced Materials and Manufacturing, The University of Tennessee, Knoxville, Tennessee 37996, USA\looseness=-1}
\author{Kipton Barros}
\affiliation{Theoretical Division and CNLS, Los Alamos National Laboratory, Los Alamos, New Mexico, 87545, USA}
\author{Richard~T.~Scalettar}
\affiliation{Department of Physics, University of California, Davis, California 95616, USA}

\date{\small\today}
\begin{abstract}
The Holstein model is a paradigmatic description of the electron-phonon interaction, in which electrons couple to local dispersionless phonon modes, independent of momentum. The model has been shown to host a variety of ordered ground states such as charge density wave (CDW) order and superconductivity on several geometries, including the square, honeycomb, and Lieb lattices. In this work, we study CDW formation in the Holstein model on the kagome lattice, using a recently developed hybrid Monte Carlo simulation method. We present evidence for $\sqrt{3} \times \sqrt{3}$ CDW order at an average electron filling of $\langle n \rangle =2/3$ per site, with an ordering wavevector at the $K$-points of the Brillouin zone. We estimate a phase transition occurring at $T_{\mathrm{c}}\approx t/18$, where $t$ is the nearest-neighbor hopping parameter. Our simulations find no signature of CDW order at other electron fillings or ordering momenta for temperatures $T \geq t/20$.
\end{abstract}
\maketitle

\section*{Introduction} \label{sec:Intro}

The interaction between electrons in a solid and the vibrations of its nuclei (phonons) can induce a variety of ordered phases \cite{Gruner1988, Gorkov1989, Zhu2015, Migdal1958, Eliashberg1960, BCS1975}. This electron-phonon coupling modifies the effective mass of itinerant electrons, and the resulting dressed quasiparticles (polarons) can pair and condense into a superconducting (SC) phase or form a periodic modulation of electron density, i.e.~CDW order. At low temperatures these various phases can compete or potentially coexist. Over the past several decades, studies of model Hamiltonians describing electron-phonon coupling have attempted to capture the interplay between their emergent ordered phases. In particular, the Holstein model \cite{Holstein1959} has been subject to much numerical and analytical study because it incorporates a simplified electron-phonon interaction into a straightforward tight-binding Hamiltonian, yet exhibits a variety of competing ordered ground states. 

A key feature of the Holstein model is an on-site momentum-independent electron-phonon coupling, which leads to an effective electron-electron attraction. Phonons are modeled as quantum harmonic oscillators of fixed frequency $\omega_0$ situated on each site of a lattice, with their motion independent of their neighbors. At low temperatures and at particular electron filling fractions, numerical studies have revealed the emergence of CDW order on square \cite{Scalettar1989, Noack1991, Vekic1992, Niyaz1993, Marsiglio1990, Costa2018, CohenStead2019, Xiao2021, Hohenadler2019, Johnston2013, Bradley2021, Esterlis2018, Dee2019, Dee2020, Nosarzewski2021}, triangular \cite{Li2019}, cubic \cite{CohenStead2020}, and honeycomb lattices \cite{Zhang2019, Feng2020_1}, with the transition temperature being sensitive to lattice geometry and dimensionality. A recent study of the Lieb lattice has also established the existence of CDW order in the Holstein model in a flat band system \cite{Feng2020_2}. 

In recent years, kagome lattices have attracted attention as a host of exotic phases owing to their high degree of geometrical frustration, and the presence of a flat band. The spin-$1/2$ kagome lattice Heisenberg antiferromagnet (KHAF) with nearest-neighbor interactions lacks any magnetic ordering, but the exact nature of the ground state been subject to much debate, with several candidates such as the Dirac spin-liquid, $Z_2$ spin-liquid, and valence bond crystal proposed \cite{Singh2007, Yan2011, Liao2017, Lauchli2019}. A recent study of the KHAF in the presence of spin-lattice coupling has shown that introducing Einstein phonons on each site can induce a magnetically ordered phase \cite{Gen2022}. For example, a $\sqrt{3} \times \sqrt{3}$ ordered phase with a $1/3$-magnetization plateau emerges in weak magnetic field, breaking a $Z_3$ symmetry, with the transition belonging to the 3-state Potts model universality class. The ordering wavevector for this phase lies at the $K$-points, i.e.~corners, of the hexagonally-shaped Brillouin zone.

The ground state properties of the half-filled kagome lattice Hubbard model are also debated. Dynamical mean field theory (DMFT) and determinant quantum Monte Carlo (DQMC) studies have identified a metal-insulator transition (MIT) in the range $U_\mathrm{c}/t \sim 7\mbox{--}9$ \cite{Ohashi2006, Ohashi2007, Kaufmann2021_1}, while variational cluster approximation (VCA) calculations estimate $U_\mathrm{c}/t \sim 4\mbox{--}5$ \cite{Higa2016}. Recent density-matrix renormalization group (DMRG) calculations find a MIT at $U_\mathrm{c}/t \sim 5.4$, along with strong spin-density wave fluctuations in the translational symmetry breaking insulating phase, signaled by an enhancement in the spin structure factor at the $K$-points of the Brillouin zone \cite{Sun2021}. CDW formation on the kagome lattice has also been observed in the extended Hubbard model. At an average electron density per site of $\langle n \rangle=2/3$ or $4/3$, or at the van Hove filling $\langle n \rangle=5/6$, several types of order have been observed \cite{Kiesel2013, Wang2013, Wen2010, Ferrari2022}, including CDW, spin density wave, and bond ordered wave states. In particular, at large $V/U$ (where $V$ is the nearest-neighbor repulsion), a CDW phase with a $\sqrt{3}\times\sqrt{3}$ supercell has been proposed for $\langle n \rangle =2/3$ and $5/6$, which has been termed CDW-III in previous studies \cite{Wen2010, Ferrari2022}. In the attractive Hubbard model, recent results \cite{Zhu2022} indicate short-ranged charge correlations at $\langle n \rangle = 2/3$ satisfying the triangle rule.

Recent experiments on kagome metals such as $A$V$_3$Sb$_5$ ($A$ = K, Rb, Cs) also motivate an understanding of CDW formation on this geometry \cite{Nguyen2022, Ortiz2019, Jiang2021, Zhao2021, Ortiz2021, Li2021, Zhou2021, Ratcliff2021, Kang2022, Xie2022, Wu2022, Kang2020, Yin2020}. In these systems, charge ordering has been observed at the $M$-points, corresponding to lattice distortions that form a star-of-David or inverse star-of-David CDW pattern. This ordering wavevector coincides with saddle points in the band structure and van Hove singularities where electronic correlations are enhanced. Theoretical studies of these materials \cite{Park2021, Ye2022, Wang2022, Denner2021, Tan2021, Christensen2021, Lin2021, Feng2021_1, Feng2021_2}, including first-principles density functional theory and mean field calculations, have corroborated these findings, where CDW ordering at the $M$-points has been observed near the van Hove filling.

Finally, kagome lattices have also been achieved in ultracold atom experiments \cite{Ruostekoski09} where they have been used to examine Bose-Einstein condensation of $^{87}$Rb \cite{Jo12}, and Rydberg atoms with large entanglement entropy and topological order \cite{Samajdar21}.

Although the Holstein coupling provides a paradigmatic model of the electron-phonon interaction, the properties of the Holstein model on the kagome lattice are not yet understood, and the possible existence of CDW order remains hitherto unexplored. 
In this work, we study the kagome lattice Holstein model using a scalable algorithm based upon hybrid Monte Carlo (HMC) sampling~\cite{CohenStead2022}, and measure the charge correlations as a function of temperature, electron density, phonon frequency, and electron-phonon coupling. We present evidence for CDW order appearing at an average electron density per site of $\langle n\rangle=2/3$, with an ordering wavevector at the $K$-points of the Brillouin zone, yielding a $\sqrt{3}\times\sqrt{3}$ supercell. Away from this filling, we find no signatures of CDW order at any ordering momenta for temperatures $T \geq t/20$.

\section*{Results}\label{sec:results}
\subsection*{Kagome lattice Holstein model}

The Holstein model describes electrons coupled to local dispersionless phonon modes in a lattice through an on-site electron-phonon interaction \cite{Holstein1959}. Its Hamiltonian is
\begin{align}\nonumber
\label{ham}
\hat{H} =&-t \sum_{\langle i,j \rangle, \sigma} \left(
\hat{c}^\dagger_{i\sigma}\hat{c}^{\phantom{\dagger}}_{j\sigma} + h.c.\right)
 - \mu \sum_{i \sigma}(\hat{n}_{i\sigma}-\tfrac{1}{2}) \\
&+ \frac{1}{2}\sum_i \hat{P}_i^2 +
\frac{\omega_0^2}{2}\sum_i\hat{X}_i^2 
+ \lambda\sum_{i \sigma} \hat{n}_{i\sigma} \hat{X}_i \,\,,
\end{align}
where $\hat{c}^\dagger_{i\sigma}$ $(\hat{c}^{\phantom{\dagger}}_{i\sigma})$ are creation
(destruction) operators for an electron at site $i$ with spin
$\sigma=\{ \uparrow \downarrow \}$, $\hat{n}_{i\sigma} =
\hat{c}^\dagger_{i\sigma}\hat{c}^{\phantom{\dagger}}_{i\sigma}$ is the electron number operator, and $\mu$ is the chemical potential, which controls the overall filling fraction. The first term describes itinerant electrons hopping between nearest-neighbor sites of the lattice, with a fixed hopping parameter $t=1$ setting the energy scale.  In the non-interacting limit, the electronic bandwidth is $W=6$ for the kagome lattice. On each site $i$ are local oscillators of fixed frequency $\omega_0$, with $\hat{X}_i$ and $\hat{P}_i$ the corresponding phonon position and momentum operators, respectively, with the phonon mass normalized to $M=1$.  The local electron density
$\hat{n}_{i\sigma}$ is coupled to the displacement
$\hat{X}_i$ through an on-site electron-phonon interaction
$\lambda$, which we report here in terms of a dimensionless parameter $\lambda_\mathrm{D} = \lambda^2/\omega_0^2
\, W$.

\begin{figure*}[htb!]
\includegraphics[width=2\columnwidth]{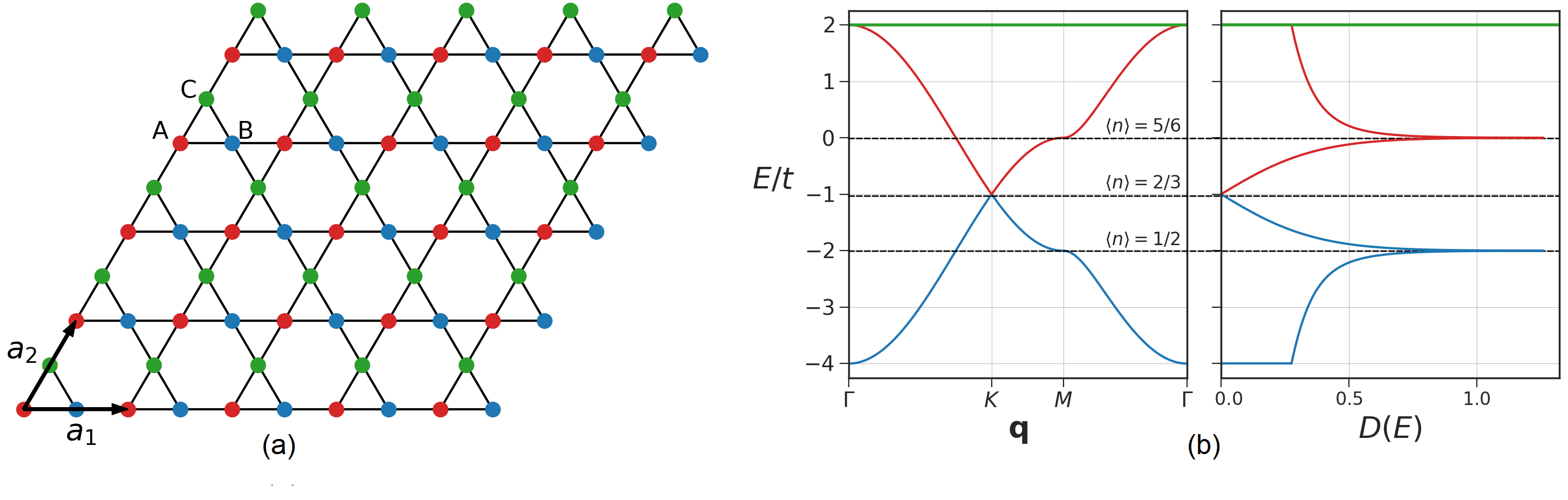}
\caption{\textbf{Kagome lattice and band structure.} (a) Geometry of the kagome lattice for $L=6$, with lattice vectors $\mathbf{a}_1 = (1,0)$ and $\mathbf{a}_2 = (\frac{1}{2}, \frac{\sqrt{3}}{2})$. Colors denote the three triangular sublattices. (b) Left: The tight-binding electronic band structure for the kagome lattice showing the three distinct bands. Dashed lines indicate the Fermi energy at specific electron densities. Right: The non-interacting density of states $D(E)$ for the kagome lattice. A delta function at $E=2t$ is due to the flat band.}
\label{Bands}
\end{figure*}

The kagome lattice vectors $\mathbf{a}_1 = (1,0)$ and $\mathbf{a}_2 = (\frac{1}{2}, \frac{\sqrt{3}}{2})$ are shown in Fig.~\ref{Bands}(a), with corresponding reciprocal lattice vectors $\mathbf{b}_1 = (2\pi, -\frac{2\pi}{\sqrt{3}})$ and $\mathbf{b}_2 = (0, \frac{4\pi}{\sqrt{3}})$, where we have set the lattice constant $a=1$. There are three sites per unit cell with basis vectors $\mathbf{u}_\mathrm{A} = (0, 0)$, $\mathbf{u}_\mathrm{B} = (\frac{1}{2}, 0)$, and $\mathbf{u}_\mathrm{C} = (\frac{1}{4}, \frac{\sqrt{3}}{4})$, forming a network of corner sharing triangles with three sublattices, as shown in Fig.~\ref{Bands}(a). Each site $i$ may instead be indexed by unit cell and the sublattice $\{A, B, C\}$, such that e.g.~$n_{\mathbf{i}, \alpha}$ denotes the electron density at the site belonging to sublattice $\alpha$ within the unit cell at position $\mathbf{i}$. In this work, we study finite size lattices with periodic boundary conditions, with linear dimension $L$ (up to $L=15$), $N=L^2$ unit cells, and $N_\mathrm{s} = 3N$ total sites. Note that discrete momentum values are given by $\mathbf{k} = \frac{m_1}{L} \mathbf{b}_1 + \frac{m_2}{L} \mathbf{b}_2$ where $m_i$ is an integer and $0 \leq m_i < L$.

There are multiple ways to break the sublattice symmetry of the kagome lattice. It is, therefore, important to construct an order parameter that will detect charge ordering independent of the charge distribution within the unit cell. 
For example, at a filling fraction of $1/3$, electrons may localize by doubly occupying only one site per unit cell, breaking a $Z_3$ symmetry. We therefore define an order parameter $\rho_\mathrm{cdw}$ that with perfect CDW order takes on one of three values $\mathrm{e}^{\mathrm{i}2\pi\left(\frac{s}{3}\right)}$, where $s=\{0,1,2\}$ corresponds to which way this symmetry is broken. The order parameter $\rho_\mathrm{cdw}$ should also be zero in the completely disordered state, where for any unit cell $\bf{i}$ we have $\langle \hat{n}_{\textbf{i}, \mathrm{A}} \rangle = \langle \hat{n}_{\textbf{i}, \mathrm{B}} \rangle = \langle \hat{n}_{\textbf{i}, \mathrm{C}} \rangle$. Hence we define
\begin{equation}
\label{rho_eq}
\rho_\mathrm{cdw} = \frac{n_\mathrm{c}}{2N} \sum_{\textbf{i}} \mathrm{e}^{-\mathrm{i} (\textbf{q}\cdot \textbf{i})} \left( \langle \hat{n}_{\textbf{i}, \mathrm{A}} \rangle + \mathrm{e}^{\mathrm{i}\frac{2\pi}{3}}\langle \hat{n}_{\textbf{i}, \mathrm{B}} \rangle + \mathrm{e}^{\mathrm{i}\frac{4\pi}{3}}\langle \hat{n}_{\textbf{i}, \mathrm{C}} \rangle \right) 
\end{equation}
where $\textbf{i}$ is a unit cell index, $N$ is the total number of unit cells, $\textbf{q}$ is the ordering wavevector, and $n_\mathrm{c}$ is a normalization constant included to fix $|\rho_\mathrm{cdw}|=1$ in the case of perfect CDW order. A structure factor that scales with system size can then be defined as $S_\mathrm{cdw}(\textbf{q}) \propto N \langle |\hat{\rho}_\mathrm{cdw}|^2 \rangle$, where again a proportionality constant can be included to fix $S_\mathrm{cdw}(\textbf{q})=N$ for the case of perfect CDW order.

For any pair of sites in the kagome lattice, we denote their density-density correlation in position space by
\begin{equation}
\label{cr_eq}
c_{\alpha, \nu}(\textbf{r}) = \frac{1}{N} \sum_{\textbf{i}} \langle \hat{n}_{\textbf{i} + \textbf{r}, \alpha} \hat{n}_{\textbf{i}, \nu} \rangle,
\end{equation}
where $\alpha$ and $\nu$ label the sublattice $\{A, B, C\}$ of the two sites, and $\textbf{r}$ is the displacement vector between their unit cells. The Fourier transform of $c_{\alpha, \nu}(\textbf{r})$ gives a generic charge structure factor
\begin{equation}
\label{S_anu_eq}
S_{\alpha, \nu}(\textbf{q}) = \sum_{\textbf{r}} e^{\mathrm{i}\textbf{q}\cdot \textbf{r}} c_{\alpha, \nu}(\textbf{r}),
\end{equation}
which provides information about the nature of an emergent CDW phase, where $\mathbf{q}$ is a discrete momentum value within the first Brillouin zone. For an ideal CDW pattern with ordering wavevector $\textbf{q}$, $S_{\alpha, \alpha}(\textbf{q})$ will reach a maximal value proportional to the number of sites, while for $\alpha \neq \nu$ the structure factor will vanish. 

In the following section, we show evidence of CDW ordering on the kagome lattice where electrons localize on only one site per unit cell but alternates cyclically between the $\{A, B, C\}$ sublattices from one unit cell to the next. To study the onset of this phase, we set $n_\mathrm{c}=1$ in Eq.~(\ref{rho_eq}) and define a charge structure factor 
\begin{align}
\label{scdw_eq}
S_\mathrm{cdw}(\textbf{q}) &= 3N \langle |\hat{\rho}_\mathrm{cdw}|^2 \rangle \nonumber\\
&= \frac{3}{4}\sum_\alpha \left( S_{\alpha, \alpha}(\textbf{q}) - \frac{1}{2} \sum_{\nu \neq \alpha} S_{\alpha, \nu}(\textbf{q}) \right). 
\end{align}
Additional details are given in Supplementary Discussion. Note that we employ a $\mu$-tuning algorithm \cite{Miles2022} to determine the chemical potential for any desired target density.

\subsection*{Measurements of charge order}

For the kagome lattice, the noninteracting tight-binding electronic structure with $t > 0$ has three separate bands, including one flat band at the highest energy ($E=2t$). The lower bands touch at two inequivalent Dirac points in the Brillouin zone, which we denote $K = (\frac{2\pi}{3}, \frac{2\pi}{\sqrt{3}})$ and $K^\prime = (\frac{4\pi}{3}, 0)$. The lower band is completely occupied at an average electron density per site of $\langle n \rangle=2/3$ (i.e.~an overall filling fraction of $f=1/3$), while the upper band is fully occupied at $\langle n \rangle=4/3$ ($f=2/3$). There are also saddle points in the band structure at the point $M=(\pi, \frac{\pi}{\sqrt{3}})$, which produce singularities in the density of states and sit at the Fermi level for average electron densities of $\langle n \rangle=1/2$ ($f=1/4$) and $\langle n \rangle=5/6$ ($f=5/12$). Fig.~\ref{Bands}(b) plots the non-interacting band structure and density of states for the kagome lattice, illustrating these features. 

To begin, we study the variation of local quantities as a function of electron density, at fixed $\omega_0$ and $\lambda_\mathrm{D}$. We set $\omega_0/t=0.1$ to facilitate CDW ordering in the Holstein model, as bipolarons should localize more readily in the limit $\omega_0/t \rightarrow 0$ due to reduced quantum fluctuations. We also fix a moderate value of the electron-phonon coupling $\lambda_\mathrm{D}=0.4$. We will discuss the rationale for this choice of parameters shortly.

\begin{figure}[t!]
\includegraphics[width=\columnwidth]{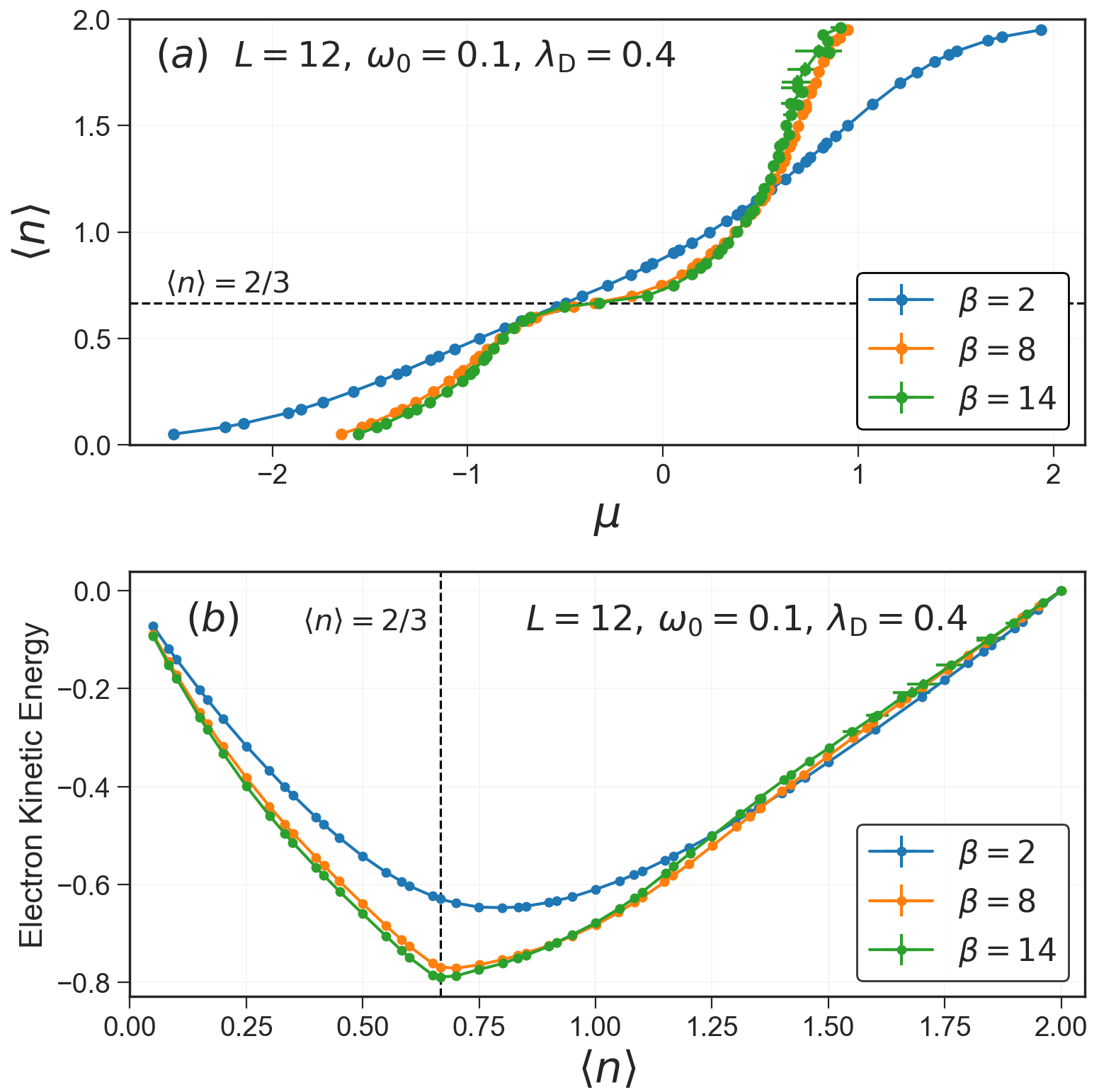}
\caption{\textbf{Electron density and kinetic energy.} (a) Average electron density per site $\langle n \rangle$ as a function of the tuned chemical potential $\mu$, for an $L=12$ lattice with $\omega_0=0.1$ and $\lambda_\mathrm{D}=0.4$ fixed. Results are shown for $\beta=2, 8,$ and $14$, with a dashed line indicating the filling $\langle n \rangle=2/3$. (b) Electron kinetic energy as a function of the electron density $\langle n \rangle$, for the same set of parameters. Error bars correspond to the standard deviation of the measured mean.}
\label{Sweep_plots}
\end{figure}

\begin{figure}[t]
\includegraphics[width=\columnwidth]{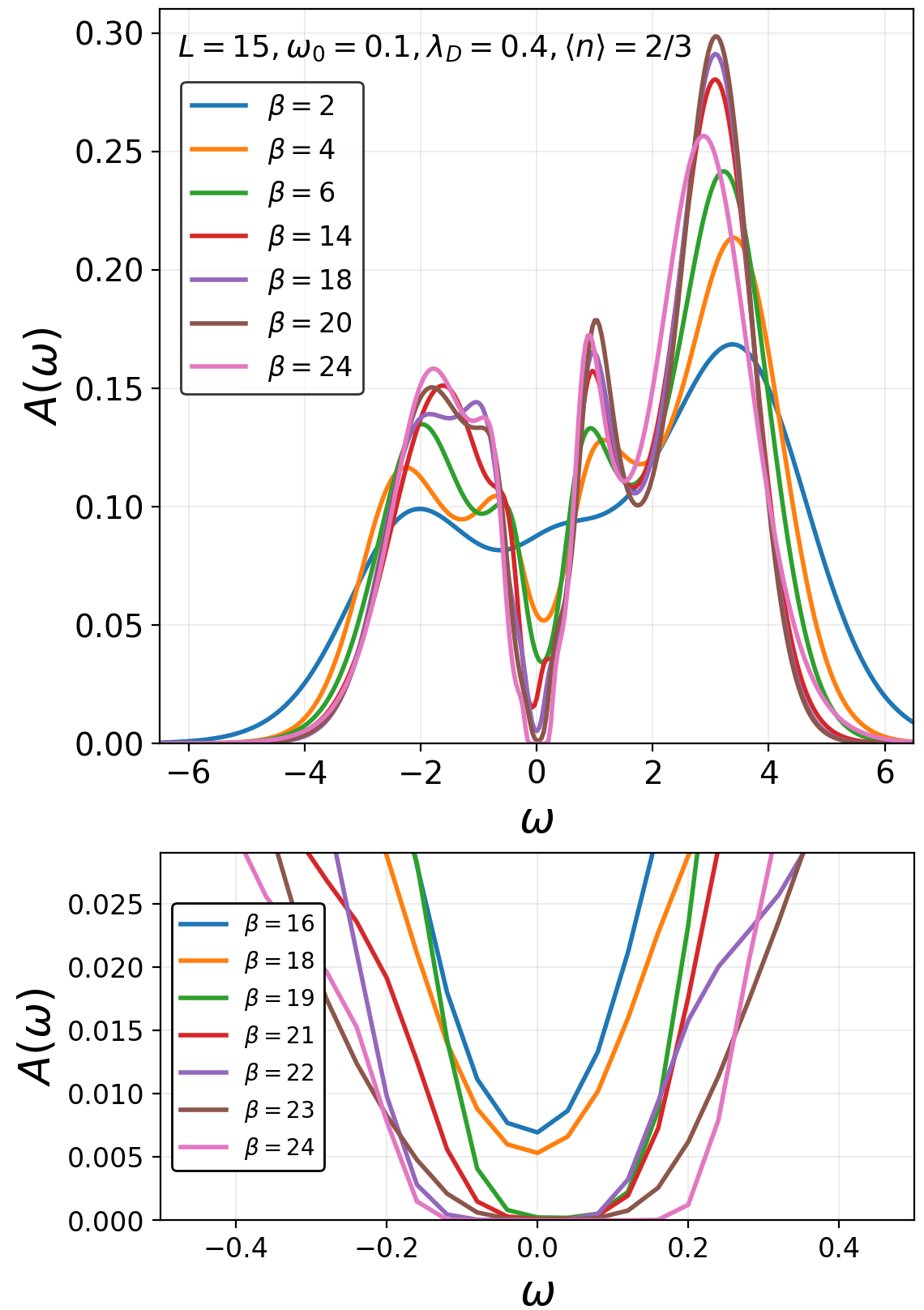}
\caption{\textbf{Spectral function.} Top: Momentum integrated spectral function $A(\omega)$ shown for a range of inverse temperatures from $\beta=2$ to $\beta=24$, at filling fraction $\langle n \rangle=2/3$ (with $\omega_0=0.1, \lambda_\mathrm{D}=0.4$). The linear lattice dimension is $L=15$ i.e.~$N_\mathrm{s}=775$. Bottom: A close-up view of the finite gap opening for $\beta \gtrsim 18$ where $A(\omega)=0$.}
\label{Spectral}
\end{figure}

In Fig.~\ref{Sweep_plots}(a) we show the average electron density per site $\langle n \rangle$ as a function of chemical potential $\mu$ for an $L=12$ lattice, as the inverse temperature is varied from $\beta=2\mbox{--}14$. We observe the formation of a plateau at $\langle n \rangle=2/3$ as the temperature is lowered, signalling the opening of a gap. No signatures of CDW ordering is observed at fillings away from $\langle n \rangle=2/3$ for these parameters. We also calculate the average electron kinetic energy as a function of electron density as shown in Fig.~\ref{Sweep_plots}(b). We observe a sharp change at $\langle n \rangle = 2/3$, where the magnitude of the electron kinetic energy becomes maximal. This is a signature of a CDW phase transition, since a configuration of doubly occupied sites surrounded by empty nearest-neighbor sites maximizes the number of bonds along which electron hopping is permitted (and corresponds to an average electron density per site $\langle n \rangle = 2/3$ on the kagome lattice). Note that since the kagome lattice is not bipartite, particle-hole symmetry is not present and thus both the kinetic energy and average filling are not symmetric about half-filling.

To further study the opening of a CDW gap as the temperature is lowered, we calculate the momentum integrated spectral function $A(\omega)$, which is related to the imaginary time dependent Green's function through the integral equation
\begin{equation}
G(\mathbf{k}, \tau) = \langle \hat{c}(\mathbf{k}, \tau) \hat{c}^\dagger(\mathbf{k}, \tau) \rangle = \int d\omega A(\mathbf{k}, \omega) \frac{\mathrm{e}^{-\omega \tau}}{1 + \mathrm{e}^{-\beta \omega}},
\end{equation}
which we invert using the maximum entropy method to obtain $A(\omega)$ \cite{Kaufmann2023}. In Fig.~\ref{Spectral}, we show the momentum integrated spectral function for an $L=15$ lattice ($N_\mathrm{s}=775$) for a range of temperatures down to $\beta=24$, again fixing $\omega_0=0.1$, $\lambda_\mathrm{D}=0.4$, and an average electron density per site of $\langle n \rangle=2/3$. We observe three peaks in the spectral function corresponding to the three-band structure. As the temperature is lowered, $A(\omega)$ reaches zero and a finite gap begins to open at $\beta \gtrsim 18$, as shown in the bottom panel, indicating a transition to an insulating CDW phase.

At an average electron density per site of $\langle n \rangle=2/3$, the lower energy band is completely filled and touches the upper band at the Dirac points $K$ and $K^\prime$. To study the onset of CDW order at this filling, we therefore calculate the charge structure factor $S_\mathrm{cdw}$ [Eq.~(\ref{scdw_eq})] evaluated at $\mathbf{q}=\mathbf{K}$, as a function of phonon frequency, electron-phonon coupling, and temperature. 

\begin{figure}[t!]
\includegraphics[width=\columnwidth]{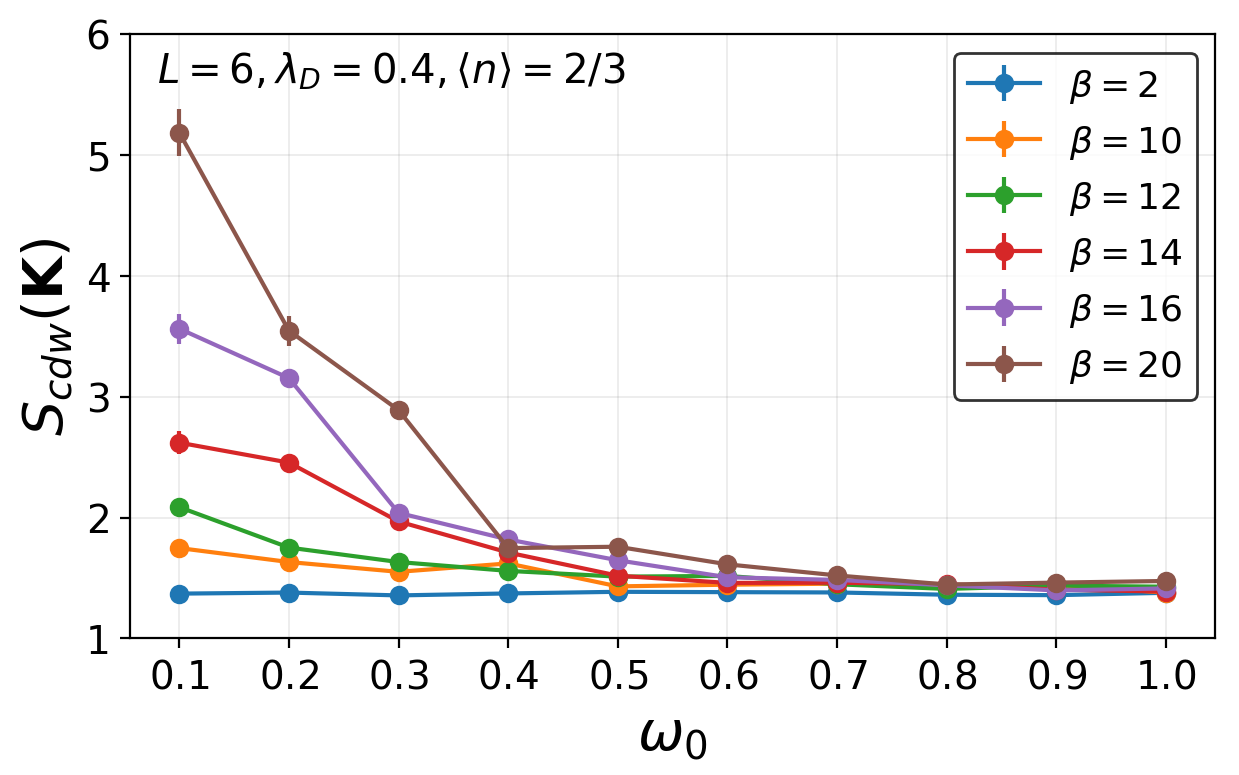}
\caption{\textbf{Charge structure factor vs.~phonon frequency.} Charge structure factor $S_\mathrm{cdw}(\mathbf{K})$ as a function of phonon frequency, for a range of temperatures from $\beta=2$ to $\beta=20$. Results are shown for an $L=6$ lattice at $\langle n \rangle=2/3, \lambda_\mathrm{D}=0.4$. Error bars correspond to the standard deviation of the measured mean.}
\label{Omega_sweep}
\end{figure}

In Fig.~\ref{Omega_sweep} we show the variation of $S_\mathrm{cdw}(\mathbf{K})$ as the phonon frequency $\omega_0$ is increased from $0.1$ to $1.0$. In the antiadiabatic limit ($\omega_0 \rightarrow \infty$), 
deformation of the lattice is weakened as sites respond more quickly to electron hopping and bipolarons do not readily localize, inhibiting the formation of a stable CDW pattern \cite{Nosarzewski2021}. In addition, quantum fluctuations are enhanced at large $\omega_0$, further suppressing CDW order \cite{Li2019}. For $\omega_0 \gtrsim 0.4$ we observe no significant growth in $S_\mathrm{cdw}(\mathbf{K})$ as the temperature is lowered from $\beta=2$ to $\beta=20$. However, for $\omega_0 \lesssim 0.3$, the structure factor begins to increase in magnitude as the temperature is reduced, growing more rapidly with $\beta$ as $\omega_0 \rightarrow 0$. We therefore fix $\omega_0=0.1$, and vary the dimensionless electron-phonon coupling $\lambda_\mathrm{D}$, in order to determine the region in which CDW order at $\langle n \rangle=2/3$ is most enhanced and subsequently estimate $T_\mathrm{c}$ for these parameters. 

\begin{figure}[t!]
\includegraphics[width=\columnwidth]{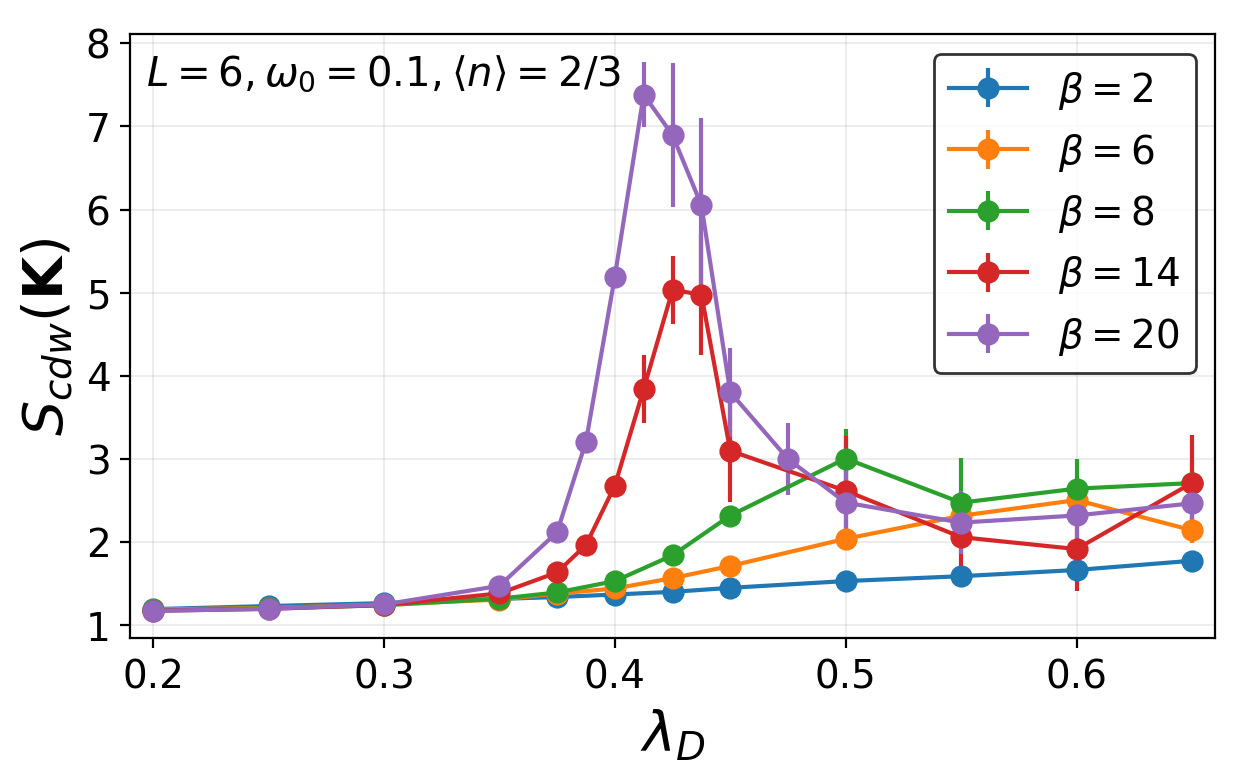}
\caption{\textbf{Charge structure factor vs.~}$\mathbf{\lambda_\mathrm{D}}$. Charge structure factor $S_\mathrm{cdw}(\mathbf{K})$ as a function of the dimensionless electron-phonon coupling $\lambda_\mathrm{D}$, for a range of temperatures from $\beta=2$ to $\beta=20$. Results are shown for an $L=6$ lattice at $\langle n \rangle=2/3, \omega_0=0.1$. Error bars correspond to the standard deviation of the measured mean.}
\label{Lambda_sweep}
\end{figure}

At small values of $\lambda_\mathrm{D}$, we find no enhancement in $S_\mathrm{cdw}(\mathbf{K})$ from $\beta=2$ to $\beta=20$ i.e.~for $\lambda_\mathrm{D} \lesssim 0.3$ there is no sign of CDW order in this temperature range, as shown in Fig.~\ref{Lambda_sweep}. This may be due to the critical temperature becoming exponentially suppressed as $\lambda_\mathrm{D} \rightarrow 0$. However, another possibility is a finite $\lambda_\mathrm{D}$ is necessary for CDW formation, as is the case in the honeycomb lattice Holstein model at half-filling \cite{Zhang2019}, which similarly has Dirac cones and a vanishing density of states at the Fermi surface. As $\lambda_\mathrm{D}$ increases, the effective electron-electron attraction is enhanced, and we observe an increase in the charge structure factor as pairs of electrons arrange themselves into a periodic CDW. As the temperature is reduced, we find that there is a maximum in $S_\mathrm{cdw}(\mathbf{K})$ at approximately $\lambda_\mathrm{D} \approx 0.4$. At larger $\lambda_\mathrm{D}$, the CDW structure factor is smaller, and eventually no significant growth is observed as the temperature is lowered from $\beta=2$ to $\beta=20$. This behavior might originate from the higher effective bipolaron mass at large $\lambda_\mathrm{D}$, which will hinder their arrangement into an ordered CDW phase, as the energy barrier associated with moving from site to site is proportional to $\lambda_\mathrm{D}$, thus promoting self-trapping. Consequently, $T_\mathrm{c}$ rapidly decreases as $\lambda_\mathrm{D}$ becomes much larger than its optimal value. We note that similar behavior has been observed in the honeycomb, square, and Lieb lattice Holstein models \cite{Zhang2019, Feng2020_2}.

\begin{figure*}[htb!]
\centering
\includegraphics[width=2\columnwidth]{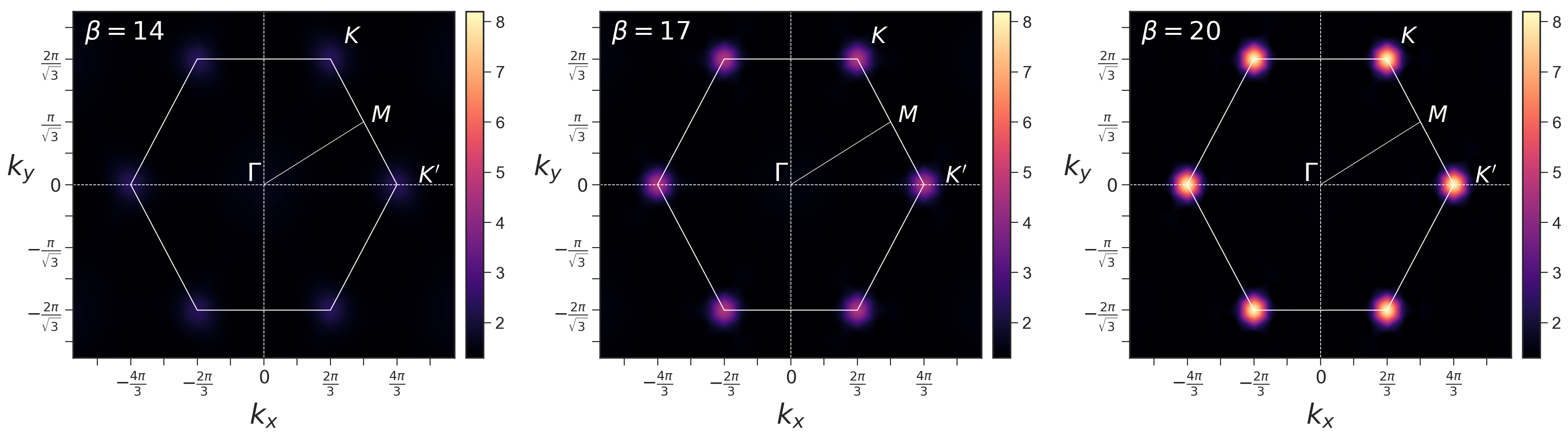}
\caption{\textbf{Charge structure factor in momentum space.} Charge structure factor $S_\mathrm{cdw}(\mathbf{q})$ shown across the Brillouin zone of the kagome lattice with $L=12$, shown for $\beta=14, 17$ and $20$. The locations of high-symmetry points in momentum space at $K = (2\pi/3, 2\pi/\sqrt{3})$, $K^\prime = (4\pi/3, 0)$, $M=(\pi, \pi/\sqrt{3})$, and $\Gamma=(0,0)$ are indicated.}
\label{BZ_Plots}
\end{figure*} 

\begin{figure*}[htb!]
\centering
\includegraphics[width=2\columnwidth]{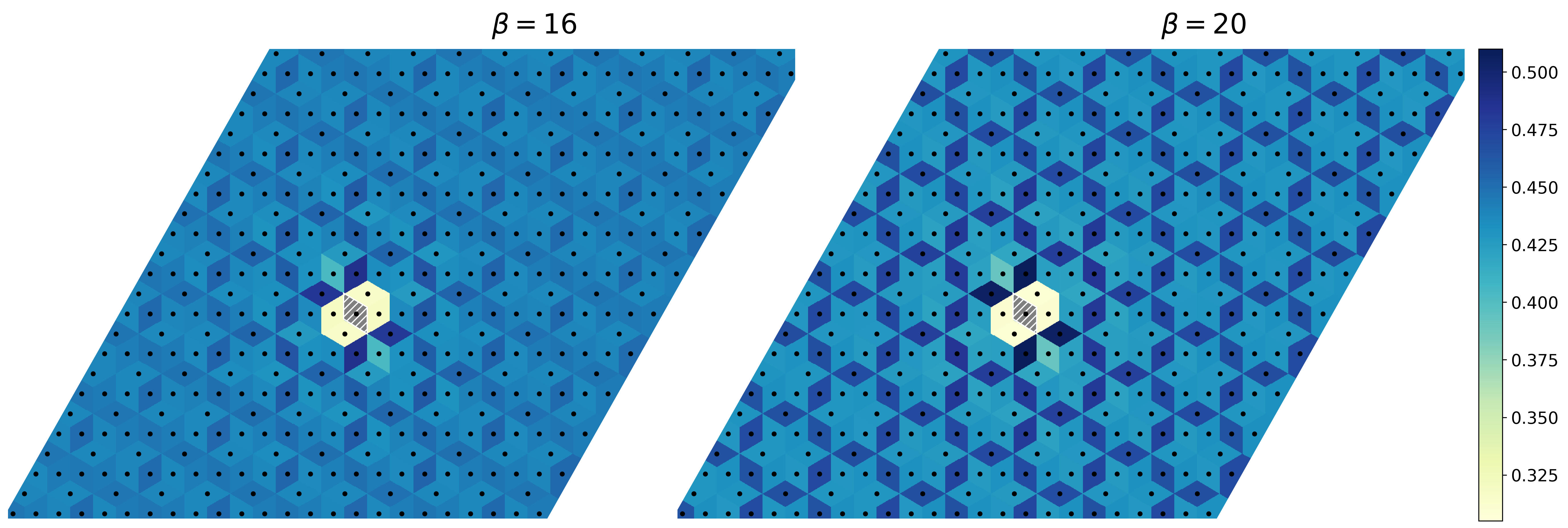}
\caption{\textbf{Real space density-density correlations.} Real space density-density correlations $\langle \hat{n}(\mathbf{0})\hat{n}(\mathbf{r})\rangle$, where $\hat{n}(\mathbf{0})$ denotes the electron density at a reference site located at the origin (gray region). For each site at position $\mathbf{r}$, the color of its Voronoi cell indicates the magnitude of $\langle \hat{n}(\mathbf{0})\hat{n}(\mathbf{r})\rangle$. Results are shown for an $L=12$ lattice with periodic boundary conditions, for $\beta=16$ (left) and $\beta=20$ (right) at filling $\langle n \rangle=2/3$ (with $\lambda_\mathrm{D}=0.4$ and $\omega=0.1$).}
\label{Den_Den_Corr}
\end{figure*}

The momentum dependence of $S_\mathrm{cdw}(\mathbf{q})$ is shown in Fig.~\ref{BZ_Plots}, where the charge structure factor at $\langle n \rangle =2/3$ is evaluated over the first Brillouin zone for an $L=12$ lattice. An enhancement in the structure factor is observed at the Dirac points as the temperature is lowered, corresponding to the onset of an ordered CDW phase, with the magnitude of $S_\mathrm{cdw}$ increasing rapidly around $\beta \gtrsim 17$. For all other momentum values, including at the $M$ and $\Gamma$-points, we find no enhancement in charge correlations with inverse temperature $\beta$, at this filling.
    
A real-space depiction of the CDW correlations at $\langle n \rangle =2/3$ is shown in Fig.~\ref{Den_Den_Corr}, which plots density-density correlations $\langle \hat{n}(\mathbf{r}) \hat{n}(\mathbf{0}) \rangle$ over an $L=12$ lattice with periodic boundary conditions. Here $\mathbf{r}=0$ is the position of a fixed reference site belonging to the $A$ sublattice. Hence Fig.~\ref{Den_Den_Corr} depicts $c_{\alpha, \nu}(\mathbf{r})$ with the origin fixed at this reference site. The CDW pattern is characterized by the localization of electron pairs on only one site per unit cell, which belongs to either the $A$, $B$, or $C$ sublattice, alternating cyclically between these from one unit cell to the next (in both the $\mathbf{a}_1$ and $\mathbf{a}_2$ directions). The fact that $\mathbf{K}$ and $\mathbf{K}^\prime$ are the ordering wavevectors for this pattern can be understood as follows. In terms of the reciprocal lattice vectors, we have $\mathbf{K}=\frac{1}{3}(\mathbf{b}_1 - \mathbf{b}_2)$ and $\mathbf{K^\prime}=\frac{1}{3}(2\mathbf{b}_1 + \mathbf{b}_2)$. If the doubly-occupied sites are separated by a displacement $\mathbf{r}=n_1 \mathbf{a}_1 + n_2 \mathbf{a}_2$, then the Fourier transform of the density-density correlation function will have peaks at $\mathbf{K}$ or $\mathbf{K^\prime}$ if $\mathbf{K} \cdot \mathbf{r} = 2m\pi$ or $\mathbf{K^\prime} \cdot \mathbf{r} = 2m\pi$, where $m \in \mathbb{Z}$. This is satisfied if $(n_1-n_2)\mod3=0$ (for $\mathbf{K}$) or $(2n_1 + n_2)\mod3=0$ (for $\mathbf{K^\prime})$, which are equivalent conditions. In other words, moving along either the $\mathbf{a}_1$ or $\mathbf{a}_2$ directions, density-density correlations will repeat with a periodicity of three unit cells, i.e.~for each unit cell, the site on which the electron pairs localize will alternate cyclically between the $\{A, B, C\}$ sublattices. For any given unit cell, the onset of this type of CDW order therefore breaks a $Z_3$ symmetry, and the phase transition should belong to the 3-state Potts model universality class. 

\begin{figure}[t!]
\centering
\includegraphics[width=\columnwidth]{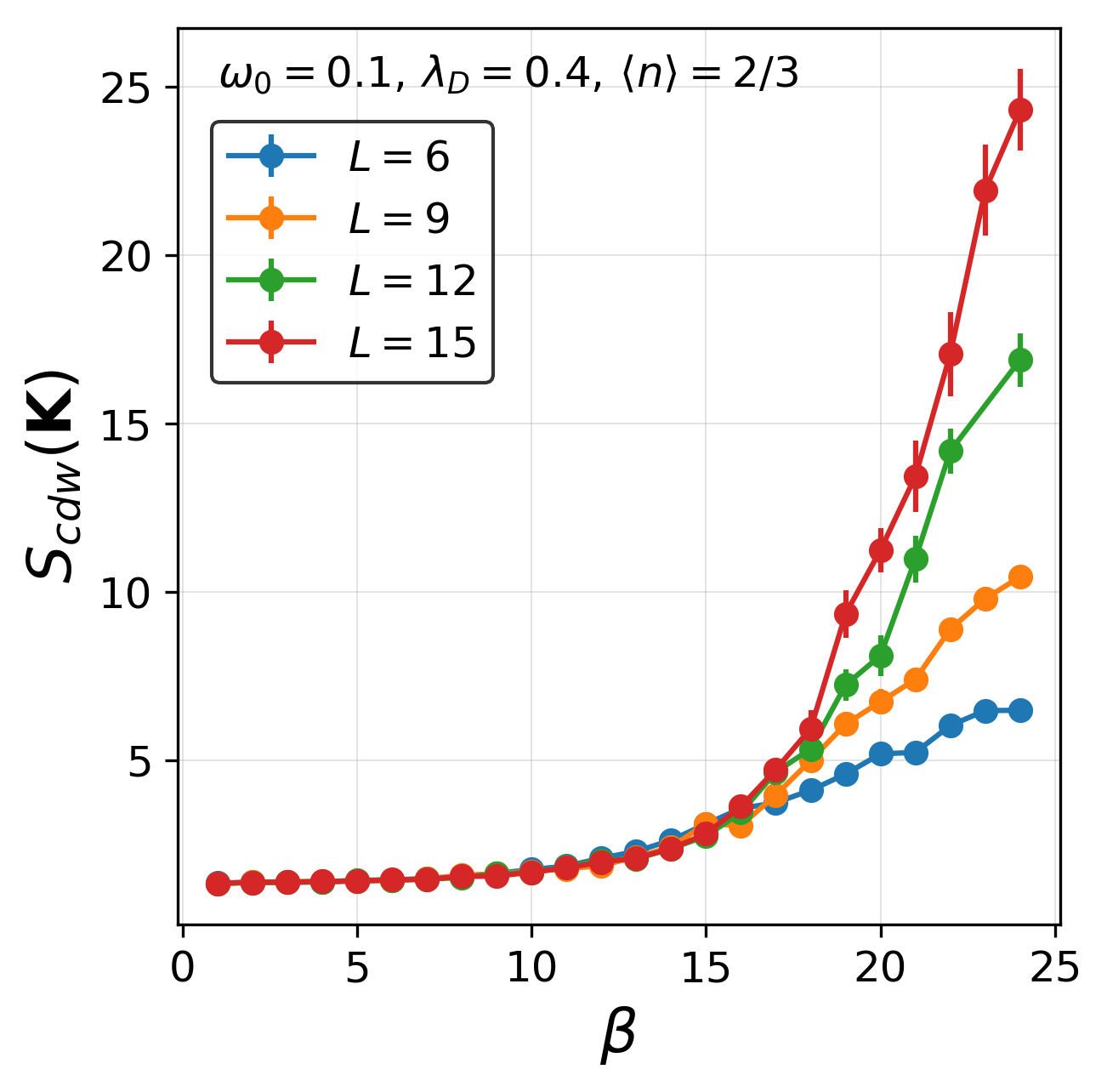}
\caption{\textbf{Charge structure factor vs.~inverse temperature.} Charge structure factor $S_\mathrm{cdw}(\mathbf{K})$ as a function of inverse temperature $\beta$, for lattice sizes $L=6, 9, 12$ and $15$, at filling $\langle n \rangle=2/3$. A lattice size dependence in the order parameter emerges at $\beta \gtrsim 18$, indicating the onset of CDW order. Here we fix $\lambda_\mathrm{D}=0.4$ and $\omega_0=0.1$. Error bars correspond to the standard deviation of the measured mean.}
\label{Beta_Sweep}
\end{figure}

\subsection*{Estimation of $T_\mathrm{c}$ for CDW phase}

For an approximate estimate of the critical temperature we can examine when the correlations become long-ranged on a finite-size lattice. As shown in Fig.~\ref{Den_Den_Corr}, at $\beta=16$ the charge order is emerging, but it is not quite long-ranged. At $\beta=20$ however, we see the periodic density correlations persist over the whole lattice. This suggests $T_\mathrm{c}$ should lie at an intermediate temperature between these two values; however, for a more accurate estimate, we must study the onset of charge order for several different lattice sizes.

\begin{figure}[t!]
\centering
\includegraphics[width=\columnwidth]{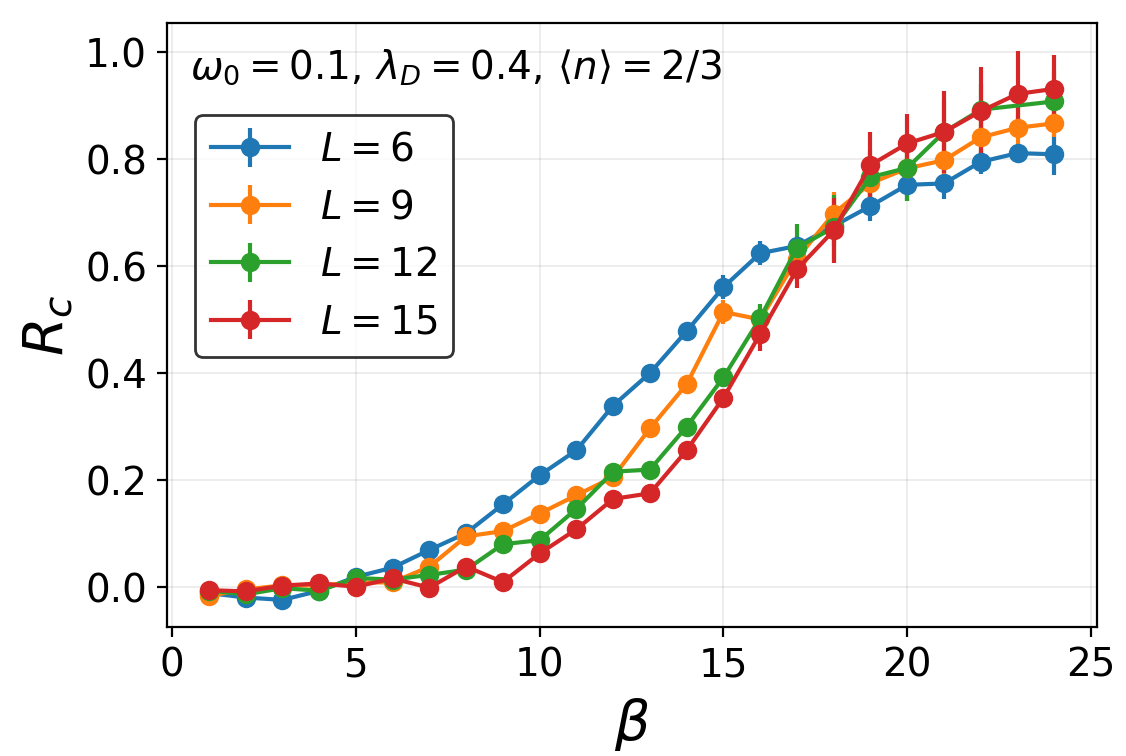}
\caption{\textbf{Correlation ratio crossing.} Correlation ratio $R_\mathrm{c}$ as a function of $\beta$, showing a crossing at $\beta_\mathrm{c} \approx 18$. Data is shown for lattice sizes $L=6, 9, 12$ and $15$, for the same parameters as in Fig.~\ref{Beta_Sweep}.}
\label{Rc_Plot}
\end{figure}

In Fig.~\ref{Beta_Sweep} we show the variation of the charge structure factor $S_\mathrm{cdw}(\mathbf{K})$ with inverse temperature $\beta$, for lattices with linear dimension $L = 6, 9, 12$ and $15$, for a range of temperatures down to $\beta=24$. At high temperatures, $S_\mathrm{cdw}(\mathbf{K})$ is relatively small and independent of lattice size. However, as the temperature is reduced, $S_\mathrm{cdw}(\mathbf{K})$ grows and becomes dependent on the lattice size for $\beta \gtrsim 18$. This signals that correlations are becoming long-ranged and thus sensitive to system size on a finite lattice, and suggests a critical temperature of $\beta_\mathrm{c} \approx 18$.  A more accurate determination of $T_\mathrm{c}$ can be made by studying the correlation ratio 
\begin{equation}
R_\mathrm{c} = 1 - \frac{S_\mathrm{cdw}(\mathbf{q} + d\mathbf{q})}{S_\mathrm{cdw}(\mathbf{q})},
\end{equation}
where the ordering wavevector $\mathbf{q}=\mathbf{K}$ here, and $|d\mathbf{q}|$ is the spacing between discrete momentum values for a lattice of linear dimension $L$. For the kagome lattice we average over the six nearest neighbors of the $K$-point in momentum space to obtain $S(\mathbf{K} + d\mathbf{q})$. The correlation ratio $R_\mathrm{c}$ is defined such that in the CDW phase, $R_\mathrm{c} \rightarrow 1$ as $L \rightarrow \infty$, (since $S_\mathrm{cdw}(\mathbf{q})$ will diverge with $L$ if there is long-range order), while $R_\mathrm{c} \rightarrow 0$ if there is no long-range order. When plotted for different lattice sizes, the crossing of $R_\mathrm{c}$ curves gives an estimate of the critical point. In Fig.~\ref{Rc_Plot} we plot $R_\mathrm{c}$ for lattices with $L=6, 9, 12$, and $15$, for the same parameters as in Fig.~\ref{Beta_Sweep} ($\langle n \rangle =2/3, \omega_0=0.1, \lambda_\mathrm{D}=0.4$). There is a crossing at $\beta_\mathrm{c} \approx 18$, which is consistent with our previous estimates of $\beta_\mathrm{c}$ obtained from observing the opening of a finite gap in $A(\omega)$, the onset of long-ranged density-density correlations, and the temperature at which $S_\mathrm{cdw}$ becomes dependent on lattice size. 

\begin{figure}[t!]
\centering
\includegraphics[width=\columnwidth]{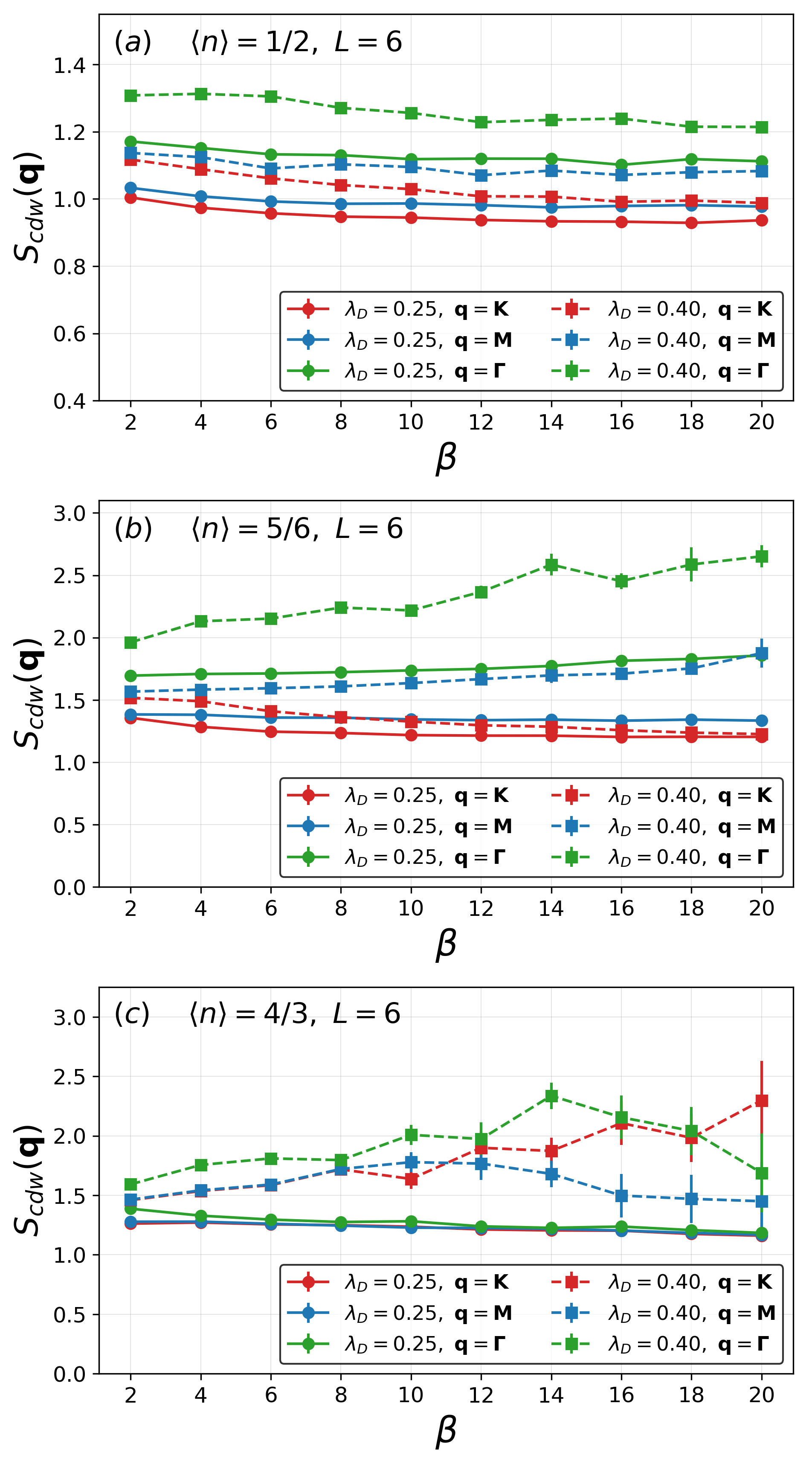}
\caption{$\mathbf{S_\mathrm{cdw}(q)}$ \textbf{at} $\mathbf{\langle n \rangle = 1/2, \, 5/6,}$ \textbf{and} $\mathbf{4/3}$\textbf{.}  Charge structure factor $S_\mathrm{cdw}(\mathbf{q})$ as a function of inverse temperature $\beta$ at several fixed electron densities: (a) $\langle n \rangle = 1/2$, (b) $\langle n \rangle = 5/6$, and (c) $\langle n \rangle = 4/3$, for an $L=6$ lattice. Data is shown for $\lambda_\mathrm{D} = 0.25$ (solid line) and $\lambda_\mathrm{D}=0.40$ (dashed line) for several momenta $\mathbf{q}$: $\mathbf{\Gamma}=(0,0)$, $\mathbf{K}=(\frac{2\pi}{3},\frac{2\pi}{\sqrt{3}})$, and $\mathbf{M}=(\pi, \frac{\pi}{\sqrt{3}}$). The phonon frequency is fixed at $\omega_0=0.1$. Error bars correspond to the standard deviation of the measured mean.}
\label{Other_densities_Sq_beta}
\end{figure}

Thus far we have studied the emergence of CDW order on the kagome lattice at a fixed electron density of $\langle n \rangle = 2/3$ per site. This choice was motivated by the observation of a CDW gap at $\langle n \rangle = 2/3$ and a sharp change in electron kinetic energy during sweeps of $\mu$ and $\langle n \rangle$, and the fact that this filling corresponds to a completely filled lower band, which meets the middle band at the Dirac points $K$ and $K^\prime$. However, we also considered fillings of $\langle n \rangle = 1/2$ and $\langle n \rangle =5/6$, i.e.~densities at which the saddle points in the non-interacting band structure (at the $M$-points) and their van Hove singularities are at the Fermi energy. 
We also consider $\langle n \rangle = 4/3$, which corresponds to completely filled lower and middle bands, with a quadratic touching at the $\Gamma$-point between the flat and middle bands (see Fig.~\ref{Bands}). In all of these cases, we find no evidence for the formation of a CDW. For example, there are no anomalous features in components of the total energy, or any indications of a plateau in the $\langle n \rangle$ vs. $\mu$ plots near these fillings, as shown in Fig.~\ref{Sweep_plots}. Moreover, as the temperature is lowered ($\beta$ increases) the charge structure factor $S_\mathrm{cdw}(\mathbf{q})$ does not grow significantly and remains relatively small in magnitude, as shown in Fig.~\ref{Other_densities_Sq_beta} for several high-symmetry points $\mathbf{q}$ in the Brillouin zone [$\mathbf{\Gamma}=(0,0)$, $\mathbf{K}=(\frac{2\pi}{3},\frac{2\pi}{\sqrt{3}})$, and $\mathbf{M}=(\pi,\frac{\pi}{\sqrt{3}})$]. We fix $\omega_0=0.1$ here to avoid suppression of any potential CDW order, which occurs in the antiadiabatic limit. These results thus suggest an absence of any charge ordering at these fillings, at least for inverse temperatures $\beta<20$. In other words, our results show no evidence for other varieties of CDW order in the kagome lattice Holstein model other than at the $K$-points at $\langle n \rangle =2/3$.

\section*{Discussion}\label{sec:discussion}

We performed hybrid Monte Carlo simulations of the Holstein model on the kagome lattice on systems of up to $N_\mathrm{s} = 775$ sites, and studied the onset of CDW order while varying the electron filling, phonon frequency, electron-phonon coupling, and temperature. Our HMC algorithm allows us to simulate larger system sizes and access lower, more realistic phonon frequencies than in previous DMQC studies of the Holstein model. We observe evidence of CDW order at an average electron density of $\langle n \rangle = 2/3$ per site (i.e.~an overall filling fraction of $f=1/3$), signaled by the opening of a gap in $A(\omega)$ at the Fermi surface, long-ranged density-density correlations, and the extensive scaling of the charge structure factor $S_\mathrm{cdw}(\mathbf{K})$ below the critical temperature. From our analysis of the correlation ration $R_\mathrm{c}$, we estimate a CDW transition at $T_\mathrm{c} \approx t/18 = W/108$, where $W$ is the non-interacting electronic bandwidth. 

This value of $T_\mathrm{c}$ is notably lower than the CDW transition temperatures found in the Holstein model on alternative geometries, e.g.~at $\lambda_\mathrm{D}=0.4$, $T_\mathrm{c} \approx t/6$ on the honeycomb and Lieb lattices, while $T_\mathrm{c} \approx t/4$ on the square lattice \cite{Zhang2019, Feng2020_2}. Moreover, the CDW order appears only for a narrow range of electron-phonon coupling strengths in the kagome lattice, peaked at $\lambda_\mathrm{D} \approx 0.4$ (for $\omega_0/t = 0.1$). In contrast, previous Holstein model studies on square, honeycomb, and Lieb lattices have found CDW transitions across a broad range $\lambda_\mathrm{D} \in [0.25, 1]$ \cite{Zhang2019, Nosarzewski2021, Feng2020_2}. On bipartite geometries with equal numbers of $A$ and $B$ sites, such as the square and honeycomb lattices, CDW formation in the Holstein model occurs at half-filling i.e.~$\langle n \rangle = 1$. However, on the Lieb lattice, for which $N_\mathrm{A} \neq N_\mathrm{B}$, when CDW order forms the density shifts away from half-filled to either $\langle n \rangle = 2/3$ or $\langle n \rangle = 4/3$, corresponding to completely filled lower and flat bands, respectively \cite{Feng2020_2}. Although the kagome lattice similarly exhibits a three-band structure, the geometry is frustrated, unlike the Lieb case, and we find that charge order emerges only at $\langle n \rangle = 2/3$ for temperatures $T \geq t/20$ with an ordering wavevector at the $K$-points and a $\sqrt{3}\times\sqrt{3}$ supercell. Our simulations did not reveal CDW order at other ordering momenta or electron densities, including at the van Hove filling. 

The CDW order we find is analogous to the $\sqrt{3}\times\sqrt{3}$ long-range magnetic order observed in the kagome lattice Heisenberg antiferromagnet, when it is coupled to local site-phonon modes \cite{Gen2022}. The same CDW phase has also been proposed as the ground state in certain regimes of the extended Hubbard model \cite{Wen2010, Ferrari2022}, i.e.~at fillings of $\langle n \rangle =2/3$ and $\langle n \rangle =5/6$ for large $V/U$, where $U$ is the on-site Hubbard term and $V$ is the nearest-neighbor repulsion, and has been termed CDW-III in these studies.

It should be noted that the CDW order we observe does not correspond to the star-of-David or inverse star-of-David patterns observed recently in kagome metals such as $A$V$_3$Sb$_5$ ($A$ = K, Rb, Cs), which exhibit ordering at the $M$-points. A recent work \cite{Ferrari2022} showed that such a CDW ordering is observed in the kagome lattice Hubbard model when a Su-Schrieffer-Heeger electron-phonon coupling is introduced. Here the electron-phonon coupling modulates the electron hopping term, and is conceptually distinct from Holstein model, in which electrons and phonons interact on a single site, rather than on the bonds of the lattice. 

\section*{Methods}\label{methods}

\subsection*{Hybrid Monte Carlo simulation}

Previous finite temperature studies of the Holstein model have typically employed DQMC~\cite{Blankenbecler1981, White1989}. In this method, the inverse temperature $\beta = L_{\mathrm{t}}\Delta\tau$ is discretized along an imaginary time axis with $L_\mathrm{t}$ intervals of length $\Delta\tau$, and the partition function is expressed as $Z = \Tr \mathrm{e}^{-\beta \hat{H}} = \Tr \mathrm{e}^{-\Delta\tau \hat{H}} \mathrm{e}^{-\Delta\tau\hat{H}} \ldots \mathrm{e}^{-\Delta\tau \hat{H}}$. Since Eq.~(\ref{ham}) is quadratic in fermionic operators, these can be traced out, giving an expression for $Z$ in terms of the product of two identical matrix determinants $\det M(x_{i, \tau})$, which are functions of the space and time-dependent phonon displacement field only. Monte Carlo sampling using local updates to the phonon field $\{x_{i,\tau} \}$ is performed and physical quantities can be measured through the fermion Green's function $G_{ij} = \langle c_{i}^\dagger c_{j} \rangle = \left[M^{-1}\right]_{ij}$. Although there is no sign problem \cite{Loh1990} for the Holstein model, these studies have been limited for two main reasons. First, the computational cost of DQMC scales as $N_\mathrm{s}^3 L^{\phantom{3}}_{\mathrm{t}}$, where $N_\mathrm{s}$ is the total number of lattice sites, prohibiting the study of large system sizes. Secondly, the restriction to local updates results in long autocorrelation times at small phonon frequencies. This aspect has limited simulations to phonon frequencies of $\omega_0 \gtrsim t$, which is unrealistic for most real materials, and is far from the regime where CDW order in the Holstein model is typically the strongest  ($\omega_0 \ll t$).

Significant efficiency gains are possible by using a dynamical sampling procedure that updates the entire phonon field at each time-step~\cite{Beyl2018, Batrouni2019}.
In this work, we use a recently developed collection of techniques to perform finite temperature simulations on extremely large clusters
\cite{CohenStead2022}. Our HMC-based approach achieves a near-linear scaling with system size \cite{Beyl2018, Duane1987, CohenStead2022_2}, allowing us to study lattices of up to $N_\mathrm{s}=775$ sites at temperatures as low as $T = t/24$. Our algorithm efficiently updates the phonon field simultaneously, allowing study of a realistic phonon frequency $\omega_0/t = 0.1$.

Near-linear scaling is achieved by rewriting each matrix determinant $\det M$ as a multi-dimensional Gaussian integral involving auxiliary fields $\Phi_\sigma$ that will also be sampled. Here, the partition function becomes
\begin{equation}
\mathcal{Z}\approx\left(2\pi\right)^{N_\mathrm{s} L_{\mathrm{t}}}\int\mathcal{D}\Phi_{\uparrow}\mathcal{D}\Phi_{\downarrow}\mathcal{D}x\,\mathrm{e}^{-S(x,\Phi_{\sigma})},
\end{equation}
where the total action is
\begin{align}
S(x,\Phi_{\sigma}) =& S_{\mathrm{B}}(x)+S_{\mathrm{F}}(x,\Phi_{\sigma})
\end{align}
with the fermionic (F) and bosonic (B) contributions
\begin{align}
\label{SF_eqn}
S_{\mathrm{F}}\left(x,\Phi_{\sigma}\right) =& \frac{1}{2}\sum_{\sigma}\Phi_{\sigma}^{T}\left(M^{T}M\right)^{-1}\Phi_{\sigma}\\
\label{SB_eqn}
S_{\mathrm{B}}(x) =& \frac{\Delta\tau}{2}\sum_{i,\tau}\left[\omega_{0}^{2}x_{i,\tau}^{2}+\left(\frac{x_{i,\tau+1}-x_{i,\tau}}{\Delta\tau}\right)^{2}\right].
\end{align}
A Gibbs sampling procedure is then adopted where $\Phi_\sigma$ and $x$ are alternately updated. The auxiliary field $\Phi_\sigma$ may be directly sampled. Using HMC, global updates to the phonon fields $x$ can be performed by introducing a conjugate momentum $p$ and evolving a fictitious Hamiltonian dynamics using a symplectic integrator \cite{CohenStead2022}.

\section*{Data Availability}
The data that support the findings of this study will be made available upon reasonable requests to the corresponding author.

\section*{Code Availability} 
The HMC code used in this study is available at \url{https://github.com/cohensbw/ElPhDynamics}.

\section*{Acknowledgements}
The work of O.B., B.C-S., S.J.~and R.T.S.~were supported by the U.S.~Department of Energy, Office of Science, Office of Basic Energy Sciences, under Award Number DE-SC0022311. K.B.~acknowledges support from the Center of Materials Theory as a part of the Computational Materials Science (CMS) program, funded by the U.S.~Department of Energy, Office of Basic Energy Sciences. 

\section*{Author Contributions}
O.B.~performed the simulations and carried out the calculations. B.C-S.~and K.B. developed the computer codes. R.T.S.~supervised the project. 
All authors discussed and analysed the results, and contributed to writing the paper.

\section*{Competing Interests}
The authors declare no competing interests.

\bibliography{new_arxiv}

\newpage

\widetext
\begin{center}
\textbf{\large Supplementary Information: Charge order in the kagome lattice Holstein model: A hybrid Monte Carlo study}
\end{center}

\vspace{10pt}

\twocolumngrid

\section*{Supplementary Discussion}
\setcounter{equation}{0}
\renewcommand{\theequation}{S\arabic{equation}}

\renewcommand{\thefigure}{S\arabic{figure}}
\setcounter{figure}{0}

\subsection*{A. Derivation of CDW order parameter}

For bipartite geometries such as the square lattice, checkerboard CDW order can occur in the Holstein model at half-filling. In these cases, electron pairs localize on one of the two sublattices ($A$ or $B$), breaking a $Z_2$ symmetry, with ordering wavevector $\mathbf{q}=(\pi,\pi)$. A charge structure factor that scales with system size can be defined as $S_\mathrm{cdw}=\sum_{\mathbf{r}} \mathrm{e}^{-\mathrm{i}(\pi, \pi)\cdot \mathbf{r}} c(\mathbf{r})$, where the sum is over all unit cells, and $c(\mathbf{r})=\frac{1}{N}\sum_{\mathbf{i}} \langle \hat{n}_{\mathbf{i}+\mathbf{r}} \hat{n}_{\mathbf{i}} \rangle$ is the real space density-density correlation function. However, one can also express $S_\mathrm{cdw}$ in terms of an order parameter $\rho_\mathrm{cdw}$ i.e.~$S_\mathrm{cdw} = N\langle |\rho_\mathrm{cdw}|^2 \rangle$, such that with perfect CDW order $\rho_\mathrm{cdw}=\pm1$ depending on which sublattice the electrons localize on, and $\rho_\mathrm{cdw}=0$ in the disordered phase. To detect checkerboard order on the square lattice no matter how the $Z_2$ symmetry is broken, we should consider the \textit{difference} between $\langle \hat{n}_\mathrm{A} \rangle$ and $\langle \hat{n}_\mathrm{B} \rangle$, i.e.~define
\begin{align}\nonumber
\rho_\mathrm{cdw} &= \langle \hat{\rho}_\mathrm{cdw} \rangle =
\frac{1}{2}\left( \langle \hat{n}_\mathrm{A} \rangle - \langle \hat{n}_\mathrm{B} \rangle \right)\\\nonumber
&= \frac{1}{2}\left( \frac{2}{N} \sum_{\mathbf{i}\in A} \langle \hat{n}_\mathbf{i} \rangle - \frac{2}{N} \sum_{\mathbf{i}\in B} \langle \hat{n}_\mathbf{i} \rangle \right)\\\nonumber
&= \frac{1}{N} \left(\sum_{\mathbf{i}\in A} \langle \hat{n}_\mathbf{i} \rangle - \sum_{\mathbf{i}\in B} \langle \hat{n}_\mathbf{i} \rangle \right) \\\nonumber
&= \frac{1}{N} \sum_{\mathbf{i}} (-1)^{i_x + i_y} \langle \hat{n}_i \rangle \\
&= \frac{1}{N} \sum_{\mathbf{i}} \mathrm{e}^{\mathrm{i}(\pi,\pi)\cdot\mathbf{i}}\langle \hat{n}_\mathbf{i} \rangle,
\end{align}
where $\mathbf{i} = i_x \hat{\mathbf{x}} + i_y \hat{\mathbf{y}}$ are the locations
of sites in a square lattice, with the lattice constant normalized to $a=1$.
Taking the squared magnitude of $\rho_\mathrm{cdw}$, a structure factor that scales with system size can then be expressed as 
\begin{align}\nonumber
S_\mathrm{cdw} &= N \langle |\hat{\rho}_\mathrm{cdw}|^2 \rangle = N \langle \hat{\rho}_\mathrm{cdw}^\dagger \hat{\rho}_\mathrm{cdw} \rangle \label{S_rho_def}\\\nonumber
&= N \Bigg\langle \bigg( \frac{1}{N} \sum_{\mathbf{j}} \mathrm{e}^{-\mathrm{i}(\pi,\pi)\cdot \mathbf{j}}  \hat{n}_\mathbf{j} \bigg)
\bigg( \frac{1}{N} \sum_{\mathbf{i}} \mathrm{e}^{\mathrm{i}(\pi,\pi)\cdot \mathbf{i}}  \hat{n}_\mathbf{i} \bigg) \Bigg\rangle\\\nonumber
&=\frac{1}{N} \sum_{\mathbf{i}, \mathbf{r}} \mathrm{e}^{-\mathrm{i}(\pi, \pi)\cdot\mathbf{r}} \langle \hat{n}_{\mathbf{i}+\mathbf{r}} \hat{n}_{\mathbf{i}} \rangle\\
&= \sum_{\mathbf{r}} \mathrm{e}^{-\mathrm{i}(\pi, \pi)\cdot\mathbf{r}}c(\mathbf{r}),
\end{align}
where $\mathbf{r}=\mathbf{j} - \mathbf{i}$, i.e.~the expression in Eq.~(\ref{S_rho_def}) is equivalent to the Fourier transform of the real space density-density correlation function $c(\mathbf{r})$ for the square lattice. 

For the kagome lattice, since each unit cell consists of three sites, we introduced a generic density-density correlation function $c_{\alpha, \nu}(\mathbf{r})$ which has Fourier transform $S_{\alpha, \nu}(\mathbf{q})$, where each lower index denotes a sublattice $A$, $B$, or $C$. In our paper we discuss a CDW pattern in which electrons localize on one site per unit cell, alternating cyclically between the $A$, $B$, and $C$ sites, as shown in Fig.~\ref{Den_Den_Corr}. Unlike checkerboard order on the square lattice, CDW order of this type can occur in three ways, breaking a $Z_3$ symmetry. As in the square lattice case, we can define an order parameter $\rho_\mathrm{cdw}$ which takes on a different value depending on how the symmetry is broken, but with $|\rho_\mathrm{cdw}|=1$ in the case of perfect order and $|\rho_\mathrm{cdw}|=0$ in the disordered phase. Since a $Z_3$ symmetry is broken, we should have $\rho_\mathrm{cdw}=\mathrm{e}^{\mathrm{i}2\pi \left(\frac{s}{3}\right)}$ (where $s=\{0,1,2\}$) in the ordered phase, and $\rho_\mathrm{cdw}=0$ in the disordered phase. However, unlike the simpler checkerboard order, the electron densities on each sublattice $\langle \hat{n}_\mathrm{A} \rangle$, $\langle \hat{n}_\mathrm{B} \rangle$, and $\langle \hat{n}_\mathrm{C} \rangle$ will vary from unit cell to unit cell, with a periodicity set by the ordering wavevector $\mathbf{q}$. We can therefore define an order parameter
\begin{equation}
\rho_\mathrm{cdw} = \frac{1}{2N} \sum_{\textbf{i}} \mathrm{e}^{-\mathrm{i} (\textbf{q}\cdot \textbf{i})} \left( \langle \hat{n}_{\textbf{i}, \mathrm{A}} \rangle + \mathrm{e}^{\mathrm{i}\frac{2\pi}{3}}\langle \hat{n}_{\textbf{i}, \mathrm{B}} \rangle + \mathrm{e}^{\mathrm{i}\frac{4\pi}{3}}\langle \hat{n}_{\textbf{i}, \mathrm{C}} \rangle \right),
\end{equation}
which satisfies these properties, where the sum is over all unit cells. As before, we can now write a structure factor that scales with system size,
\begin{align}
S_\mathrm{cdw}(\mathbf{q}) = 3N \langle |\hat{\rho}_\mathrm{cdw}|^2 \rangle = 3N \langle \hat{\rho}_\mathrm{cdw}^\dagger \hat{\rho}_\mathrm{cdw}^{\phantom{\dagger}} \rangle,
\end{align}
where a constant factor has been inserted to ensure $S_\mathrm{cdw}=N$ in the case of perfect CDW order. We can now expand the above equation to obtain an expression for $S_\mathrm{cdw}$ in terms of the generic structure factors $S_{\alpha, \nu}$, which yields
\begin{align}\nonumber
S_\mathrm{cdw}(\mathbf{q}) &= \frac{3}{4N} \sum_{\mathbf{r}} \bigg[ \mathrm{e}^{\mathrm{i} \mathbf{q} \cdot \mathbf{r}} \sum_{\alpha} \bigg( c_{\alpha,\alpha}(\mathbf{r}) -\frac{1}{2} \sum_{\alpha \neq \nu} c_{\alpha,\nu}(\mathbf{r}) \bigg)\bigg]\\
&= \frac{3}{4} \sum_\alpha \bigg( S_{\alpha,\alpha}(\mathbf{q}) - \frac{1}{2} \sum_{\alpha \neq \nu} S_{\alpha, \nu}(\mathbf{q}) \bigg) \label{S_eqn_appendix},
\end{align}
where $c_{\alpha, \nu}(\mathbf{r})$ is the generic real space density-density correlation function. This expression for $S_\mathrm{cdw}$ is the measure we use to detect CDW ordering in the kagome lattice, and is the quantity we show in Fig.~\ref{Beta_Sweep} at $\langle n \rangle = 2/3$ as a function of $\beta$ for different lattice sizes. For a particular CDW pattern on the kagome lattice under study, Eq.~(\ref{S_eqn_appendix}) can be multiplied by a constant factor (if necessary) to fix $S_\mathrm{cdw}=N$ in the case of perfect CDW order.

\begin{figure}[t!]
\includegraphics[width=\columnwidth]{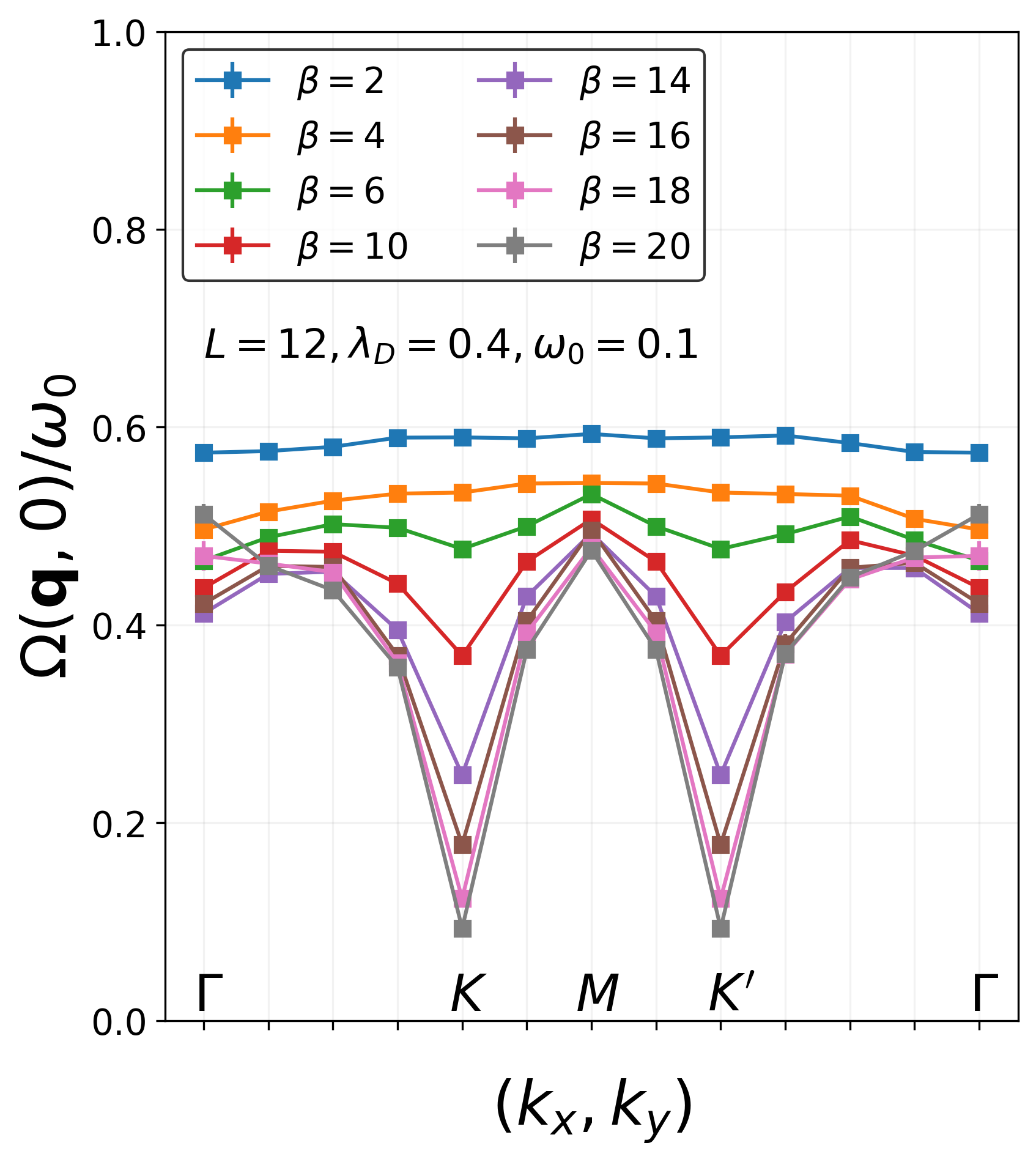}
\caption{\textbf{Renormalized phonon frequency.} The renormalized phonon frequency $\Omega(\mathbf{q}, 0)/\omega_0$ at $\langle n \rangle = 2/3$ and $\lambda_\mathrm{D}=0.4$, shown for a range of temperatures from $\beta=2$ to $\beta=20$. Phonon softening at the ordering wavevectors $K$ and $K^\prime$ is observed as the temperature is lowered.}
\label{renormalized_phonon}
\end{figure}

\subsection*{B. The renormalized phonon energy $\Omega(\mathbf{q}, \mathrm{i}\nu_n=0)$}

An additional signature of the CDW transition can be observed in the renormalized phonon energy $\Omega(\mathbf{q}, \mathrm{i}\nu_n=0)$, where a softening of the phonon dispersion is expected to occur at the ordering wavevector $\mathbf{q}_{\mathrm{cdw}}$ as the temperature is lowered. The renormalized phonon energy is given by $\Omega(\mathbf{q}, 0) = [\omega_0^2 + \Pi(\mathbf{q}, 0)]^{1/2}$, where $\Pi(\mathbf{q}, 0)$ is a function defined in terms of the momentum-space phonon Green's function $D(\mathbf{q}, \nu_n)$, through the relation
\begin{equation}
D(\mathbf{q}, \nu_n) = \frac{2\omega_0}{(\mathrm{i} \nu_n)^2 - \omega_0^2 - \Pi(\mathbf{q}, \nu_n)},
\end{equation}
where $\nu_n = 2\pi nT$ (and we set $\hbar=1$). In Fig.~\ref{renormalized_phonon} we show $\Omega(\mathbf{q}, 0) / \omega_0$ along a closed triangular path $\Gamma$--$K$--$K^\prime$--$\Gamma$ within the Brillouin zone. Results are shown for an $L=12$ lattice with electron-phonon coupling $\lambda_\mathrm{D} = 0.4$ and bare phonon frequency $\omega_0=0.1$, at a filling of $\langle n \rangle = 2/3$, where we previously observed evidence of CDW ordering.

At high temperature ($\beta \ll \beta_\mathrm{c}$) we find that the renormalized phonon dispersion is relatively flat, e.g.~for $\beta=2$, we have $\Omega(\mathbf{q}, 0)/\omega_0 \approx 0.6$ for all momenta $\mathbf{q}$. This is indicative of strong electron-phonon coupling, where the renormalized phonon frequency is uniformly suppressed relative to the bare phonon frequency even at temperatures well above $T_\mathrm{c}$. This signature of strong coupling is expected behavior in the Holstein model, and has been observed in the square lattice at similarly large values of $\lambda_\mathrm{D}$ \cite{Esterlis2018}. As the temperature is lowered, we observe a softening of the phonon dispersion at the expected ordering wavevectors $K$ and $K^\prime$. We observe sharp dips in the phonon dispersion as the temperature is reduced to $\beta \approx 18$, consistent with our estimate of $\beta_\mathrm{c}$ obtained from the crossing of $R_\mathrm{c}$ curves shown in Fig.~\ref{Rc_Plot}. Due to an expected finite-size effect, $\Omega(\mathbf{q}, 0)$ does not reach exactly zero below $T_\mathrm{c}$ in our simulations \cite{CohenStead2022_2}. \newline

\subsection*{C. HMC simulation parameters}

Here we provide additional details of the parameter values used in our HMC simulations, which are explained further in Ref.~\cite{CohenStead2022}. We fixed the imaginary-time discretization at $\Delta\tau = 0.05$,
and performed updates to the phonon field using HMC trajectories typically of $N_\mathrm{t}=50$ time-steps of size $\Delta t=0.2$. We initially thermalize
our systems by performing $N_{\mathrm{therm}}=2000\mbox{--}3000$ trial updates. This is followed by $N_{\mathrm{sim}}=2500\mbox{--}4000$ updates, where measurements are taken at the end of each trajectory. Each time-step we numerically solve for the fictitious force $\dot{p}=-\frac{\partial S}{\partial x}$ using the conjugate gradient method with a relative residual error threshold $\epsilon_{\mathrm{max}}=10^{-5}$. 

The forces in our HMC dynamics can be separated into fermionic and bosonic components $-\frac{\partial S_\mathrm{F}}{\partial x}$ and $-\frac{\partial S_\mathrm{B}}{\partial x}$ [see Eqs.~(\ref{SF_eqn}) and (\ref{SB_eqn})], where the bosonic part is much less expensive to calculate. We thus employ time-step splitting where trajectories evolve with time-step $\Delta t^\prime = \Delta t / n_\mathrm{t}$ with $n_\mathrm{t}=10$ using the bosonic force alone, followed by a single step of size $\Delta t$ using the fermionic force. We also use Fourier acceleration via a dynamical mass matrix with regularization parameter $m_{\mathrm{reg}} = \omega_0$ to further reduce autocorrelation times.


\end{document}